Preprint vs 1

# New insights into crystallographic relation and lattice dynamics effects in {CdO/MgO} superlattices grown by plasma-assisted molecular beam epitaxy

A. Wierzbicka[1], E. Przezdziecka[1], I. Perlikowski[2], E. Zielony[2], A. Adhikari[1], A. Lysak[1]

[1] Institute of Physics, Polish Academy of Sciences, Al. Lotnikow 32/46, Warsaw 02-668, Poland

[2] Department of Quantum Technologies, Wroclaw University of Science and Technology, Wybrzeze Wyspianskiego 27, 50-370 Wroclaw, Poland

**Abstract**

This article explores the structural properties of molecular beam epitaxy grown {CdO/MgO} superlattices on sapphire substrates of different crystallographic orientations (*a*-, *c*-, *r*-, and *m*-plane). The investigations involve a comprehensive analysis using X-ray diffraction and Raman spectroscopy. High-resolution X-ray diffraction studies unveil a significant influence of surface symmetry on both the substrates and the epitaxial layers, particularly with respect to the occurrence of twins in the superlattices. Remarkably, no twins are observed on r-oriented sapphire substrates, resulting in improved interface and crystallographic quality. The results of studies demonstrated in this work show that the growth rate of CdO sublayers within {CdO/MgO} superlattices is intricately dependent on the substrate orientation. Notably, the *c*-plane and *m*-plane sapphire substrates yielded thicker CdO sublayers, indicating comparable growth rates for these crystallographic orientations. Conversely, the *a*-plane and *r*-plane orientations seemed to favor a slower growth rate of CdO sublayers.

**Introduction**

Modern devices developed on new semiconductor systems require the synthesis of high-quality materials and understanding of the influence of substrates on the structure's quality and its properties. A new material with interesting properties that can be used in the development of innovative optoelectronic devices is CdMgO in which, by changing the alloy composition, it is possible to control the energy gap in the range of about 2 to 7 eV [1], [2]. Through the use of epitaxial methods, we can produce CdMgO random alloys as well as quasi-ternary {CdO/MgO} alloys for short-period superlattices (SLs) [3]. Both the aforementioned random and quasi-ternary alloys may be included in a broad class of transparent conductive oxides materials (TCOs). TCOs are widely used in various optoelectronic applications, such as solar cells, displays, touch panels and transparent electrodes [4], [5].

Heteroepitaxy is a specific type of epitaxy in which a thin film is grown on a substrate with a different crystal structure or lattice parameters, however, the growth direction of the film is oriented and aligned with the substrate [6], [7]. In the case of {CdO/MgO} SLs on sapphire, heteroepitaxy refers to the growth of CdO and MgO layers on sapphire substrates with different crystal structures. This type of growth plays a crucial role in device preparation for several reasons. First, heteroepitaxy allows for the selection of materials with compatible lattice parameters. When the lattice constants of the film and substrate are closely matched, it reduces strain and defects at the interface. This enables the growth of high-quality film and minimizes lattice mismatch-induced defects, such as dislocations, which can degrade device performance. Next, by combining different materials in a heteroepitaxial structure, it is possible to engineer a system with unique properties what is not easy to achieve using homoepitaxy [8]. For example, {CdO/MgO} SLs may exhibit enhanced electrical conductivity or optical properties by controlling the particular layers thicknesses. Finally, heteroepitaxy allows for precise control of the band-gap alignment at the interface between the film and the substrate. This is important for electronic- as well as optoelectronic devices where the energy level alignment between different materials affects the device's performance. By careful selection and controlled growth of heteroepitaxial layers, it is possible to produce structures of well-defined energy band-gaps, efficient charge carrier transport and desired device operation.

{CdO/MgO} superlattices grown by molecular beam epitaxy (MBE) on sapphire substrates of different crystal orientations (a-plane, c-plane, m-plane, r-plane) hold significant promise for a variety of applications, including next-generation electronic and optoelectronic devices. In the case of {CdO/MgO} SLs on sapphire substrates, each sapphire orientation offers distinct crystallographic properties and surface characteristics. Here are the key points regarding the importance of heteroepitaxial growth in this context. First important point is strain engineering. Heteroepitaxial growth introduces strain in the superlattice structure due to the lattice mismatch between the film and the substrate. By selecting different sapphire orientations, the strain characteristics can be tailored, offering opportunities for strain reduction. Strain can influence the electronic, optical, and magnetic properties of the superlattice, enabling the tuning of device's performance. The next important point is the interface engineering. The interfaces between different layers within the superlattice can have a significant impact on the device performance. By choosing specific sapphire orientations, it becomes possible to optimize the interface quality, which can affect the electrical conductivity, carrier mobility, and other relevant device properties. Choosing the proper substrate is a very important aspect in the production of efficient devices. Sapphire is a popular substrate material due to its excellent thermal and structural stability. The availability of sapphire substrates of different crystal orientations provides powerful possibilities for customizing the properties of epitaxial thin films and devices. The last, but not the least important thing is achieving the device integration and its compatibility. Heteroepitaxial growth of {CdO/MgO} superlattices on sapphire offers compatibility with various device integration processes. Sapphire substrates are widely used in the semiconductor industry, and different orientations provide diverse options for device design and fabrication. The ability to grow superlattices on multiple sapphire orientations expands the range of possible device structures and enables integration with existing technologies.

In this work we use two basic techniques X-ray diffraction (XRD) and Raman spectroscopy to investigate {CdO/MgO} SL structures on sapphire substrates of different crystallographic orientations. The combination of the aforementioned experimental techniques provides a powerful tool used to characterize the structural properties of materials, including thin films and superlattices. When applied to MBE-grown {CdO/MgO} superlattices, the techniques can deliver valuable information on the crystalline structure and composition of the layers within the superlattice.

**Experimental**

{CdO/MgO}SL samples were grown by plasma-assisted molecular beam epitaxy (PA-MBE) in a Compact 21 machine, on *a*-, *c*-, *r*- and *m*-plane sapphire substrates that were loaded into the reactor after chemical cleaning ($H_2SO_4:H_2O_2$). In loading chamber the substrates were thermally annealed at 150°C for 30 minutes. Prior to the growth, the sapphire substrates were chemically cleaned and then annealed at a temperature of about 700°C in a high vacuum for 30 min and finally in oxygen plasma also at 700°C. The SLs films were grown at 360°C. During the growth process the oxygen plasma parameters, such as the power and oxygen flow, were fixed as follows: 450 W and 3 ml/min. We obtained 4 types of samples which were {CdO/MgO} superlattices on *a*- (Sample A), *c*- (Sample C), *m*- (Sample M) and *r*-plane (Sample R) sapphire substrates, grown in the same growth conditions. The SL period was repeated by 17 times keeping the same MgO and CdO layers thicknesses which were of 2 nm for MgO and 4 nm for CdO. Figure 1 shows the schematic drawing of the obtained structures.

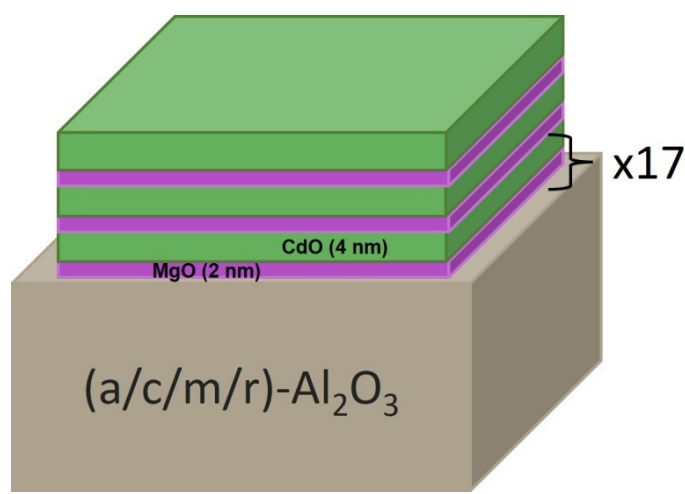

**Figure 1.** Schematic drawing of {CdO/MgO} superlattices obtained on *a*-, *c*-, *m*- and *r*-oriented sapphire substrates.

The SL samples were investigated by X-ray diffraction using X'Pert Pro MRD diffractometer. We use CuK$_{\alpha 1}$ radiation, hybrid 2-bounce monochromator (with X-ray mirror and Ge (220) crystal) and Pixcel detector.

Raman spectra measurements were performed at room temperature using the T64000 Horiba Jobin-Yvon spectrometer configured in a backscattering geometry, offering the spectral resolution in the order of 0.5 cm$^{-1}$. As a detector a liquid nitrogen-cooled multichannel silicon CCD camera was used. A 514.5 nm Ar$^+$ laser was used to excite the samples (non-resonant excitation). The measurements were performed without the detection of polarization of the scattered light.

**Results and discussion**

In order to make the article more transparent it has been divided into two parts. The first one concerns the results of X-ray diffraction studies and analysis, whereas the second part is based on Raman spectroscopy results.

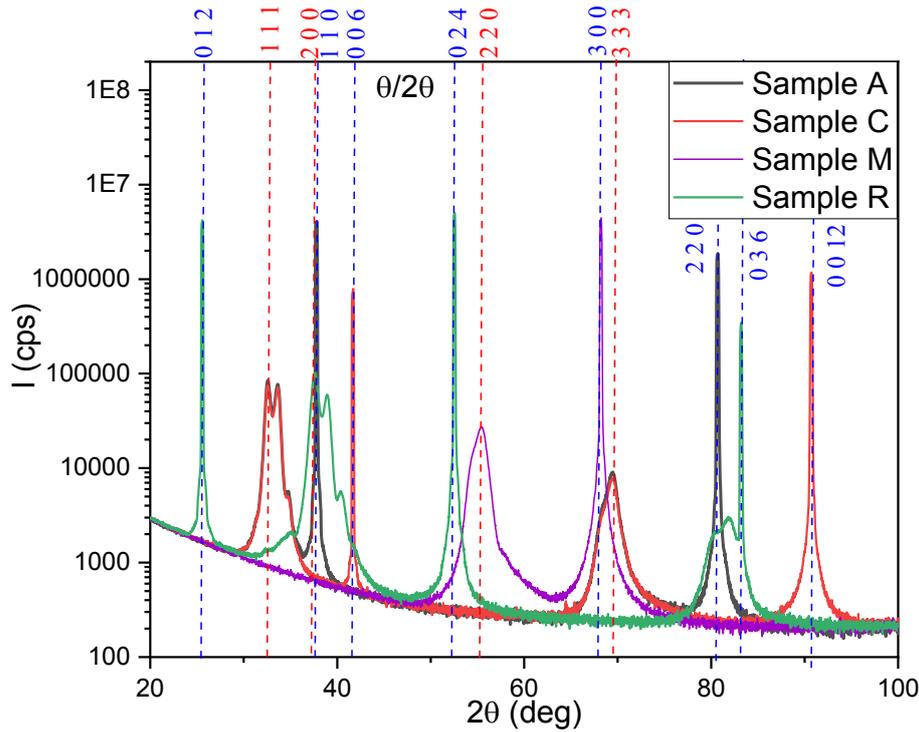

**Figure 2.** θ/2θ X-ray diffraction patterns for {CdO/MgO} SLs grown on *a*-, *c*-, *m*- and *r*-plane sapphire substrates.

1. XRD analysis

First, {CdO/MgO} SLs were investigated using a low resolution XRD system. A fully opened detector registered signal from wide parts of a given sample. XRD results are shown in Figure 2. The blue spotted lines indicate signals from sapphire substrates with different orientations: (11-20) *a*-plane, (0001) *c*-plane, (10-10) *m*-plane and (01-12) *r*-plane. The substrate was grown in a rhombohedral crystal structure. These measurements reveal the SL structure is cubic. It has been found that the SL growth direction is different depending on the substrate orientation. On *a*- and *c*-plane sapphire the {CdO/MgO} SLs grow in (111) crystallographic direction, whereas, on m-plane sapphire the SL changes growth direction to (110). For *r*-plane substrate SL grows in (100) crystallographic direction. It is consistent with scientific literature, which reports that epitaxial growth of cubic layers on sapphire substrates with different orientations behaves in the same way, choosing the most favored crystallographic direction ([2], [3], [9]). The next characteristic feature of SL system is the quality of the cubic structure of SLs. For a thorough analysis of crystallographic structure, strain distribution and defect distribution in SLs we have adopted high resolution X-ray diffraction (HRXRD).

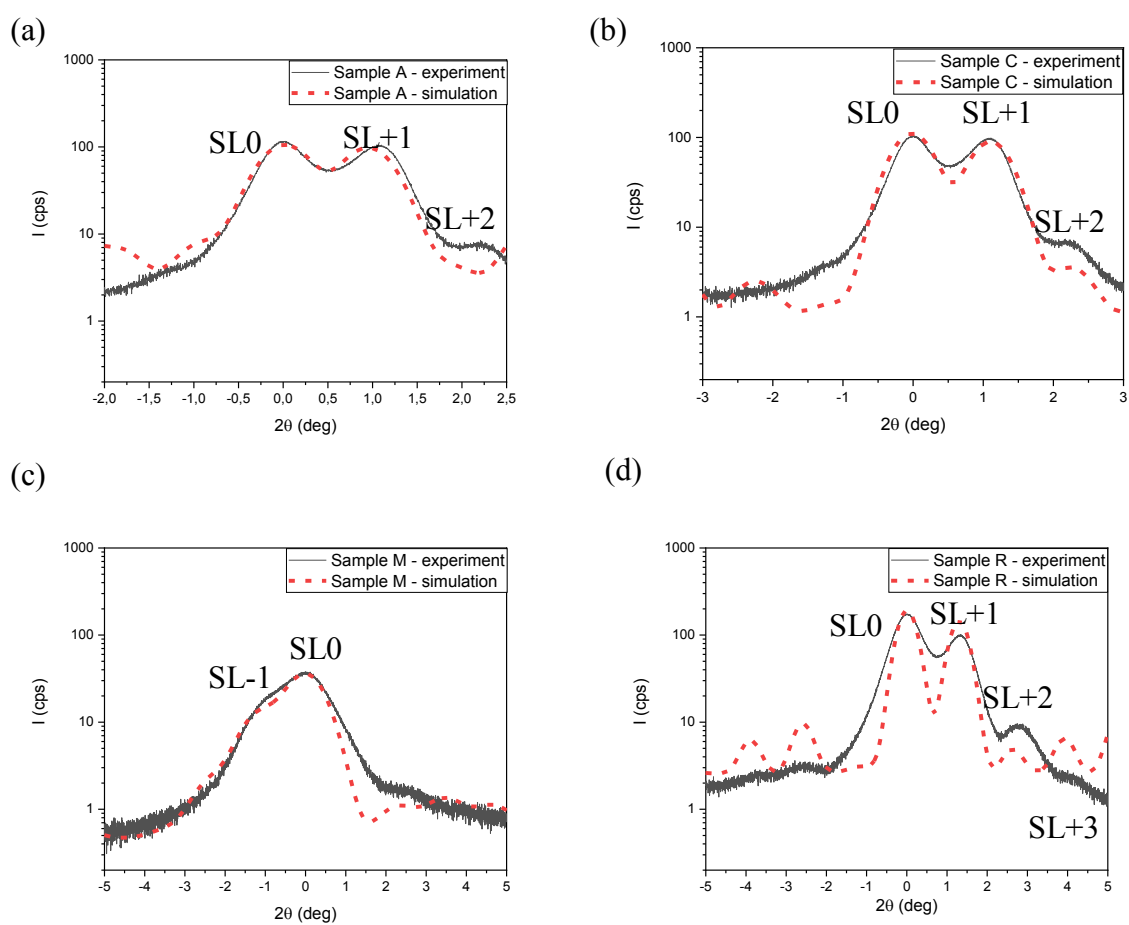

**Figure 3.** HRXRD $2\theta/\omega$ scans of {CdO/MgO} SL structures on *a*-plane (a), *c*-plane (b), *m*-plane (c) and *r*-plane (d) sapphire substrates. Solid lines show the experimental data and the dashed lines the simulation results.

HRXRD measurements were performed with a Ge (220) analyzer crystal in front of the detector. The divergence of the X-ray beam with this analyzer is of 12". Therefore, it is a great possibility to grasp residual effects present in investigated samples. First HRXRD $2\theta/\omega$ scans were measured (Figure 3) to calculate the periodicity and thickness of individual layers in {CdO/MgO} superlattices. Fringes visible on HRXRD scans are evidence of the quality of the SL. The more fringes in HRXRD $2\theta/\omega$ scans the better crystallographic quality of SL is. Consequently, for samples A, C and R we observe characteristic fringes in XRD signals. The $2\theta/\omega$ scans of sample R yield the third order of SL fringe (SL +3). Whereas, for sample M there is a broad peak coming from the zero order peak of SL indicating that for this orientation of the sapphire substrate, the growth is more difficult to implement. In our previous works (Refs. [3], [10]), the growth of {CdO/MgO} SLs on *r*-plane and *c*-plane sapphire substrate was similar and we observed several fringes on XRD scans. Here we compare for the first time results obtained for {CdO/MgO} SL grown on *a*-plane and *m*-plane sapphire. As we can see the growth in these directions is associated with deterioration of crystallographic structure of {CdO/MgO} SLs. In the next step we performed simulations of $2\theta/\omega$ profiles using X'Pert Epitaxy software. A detailed description of this software can be found elsewhere [11]. The results of the fittings are presented as dashed red lines in Figure 3. They follow the experimental data. The obtained thickness values of individual layers of the studied SLs are presented in Table 1.

| SAMPLE | $t_{CdO}$ [nm] | $t_{MgO}$ [nm] | $a_{(SL0)}$ [Å] |
|---|---|---|---|
| Sample A | 2.75 | 2.15 | 4.7479 |
| Sample C | 6.05 | 2.15 | 4.7471 |
| Sample M | 6.03 | 1.75 | 4.6774 |
| Sample R | 2.3 | 2.35 | 4.7715 |

**Table 1.** Average values of thicknesses of individual CdO ($t_{CdO}$) and MgO ($t_{MgO}$) sublayers together with lattice parameters $a_{(SL0)}$ of {CdO/MgO} SLs structures grown on sapphire substrates of different crystallographic orientation.

Analyzing the obtained values of $t_{CdO}$ and $t_{MgO}$ for CdO and MgO sublayers, one can see that they differ from each other although the growth conditions were kept the same for all the samples (it was the same growth process). Thus, it is evidence that the growth rate strongly depends on the substrate orientation. For {CdO/MgO} SLs grown on c-plane and m-plane sapphire substrates, we observe thicker CdO sublayers. It means that the growth rate is comparable for these two substrate orientations. In contrast, for sample A and R the a-plane and r-plane orientation of the sapphire substrate favors slower growth rate of CdO sublayers. For the applied growth parameters, the growth of MgO sublayers seems to be independent of the substrate orientation.

The XRD studies were extended with the measurements of HRXRD maps near the main reciprocal lattice point of the {CdO/MgO} SL. Such experiments allow to find the accurate values of lattice parameters and check the deformation of the unit cell of SLs. Figure 4 shows HRXRD reciprocal space maps measured for symmetrical 111 {CdO/MgO} reflection for sample A and sample C (Fig. 4a and Fig. 4b), 220 {CdO/MgO} reflection for sample M (Fig. 4c) and 200 {CdO/MgO} reflection for sample R (Fig. 4d). The obtained reciprocal space maps show that for samples A, C and R the SL fringes are clearly visible. Whereas, for sample M only broad two peaks of SL can be distinguished which is in agreement with the result in Fig. 3c. The measurements of HRXRD reciprocal space maps allow to calculate the accurate values of *a(SL0)* lattice parameters [12], [13] of SLs collected in Table 1. The values derived for samples A and C are similar what is in good agreement with the obtained thicknesses values of MgO and CdO in these SL structures. Comparing the calculated average values of the SL lattice parameter it can be noticed that for samples A, C and R, are close to each other. However, the lattice constant for sample M differs significantly from the other ones. This may point to lattice distortion of {CdO/MgO} SL grown on *m*-plane sapphire.

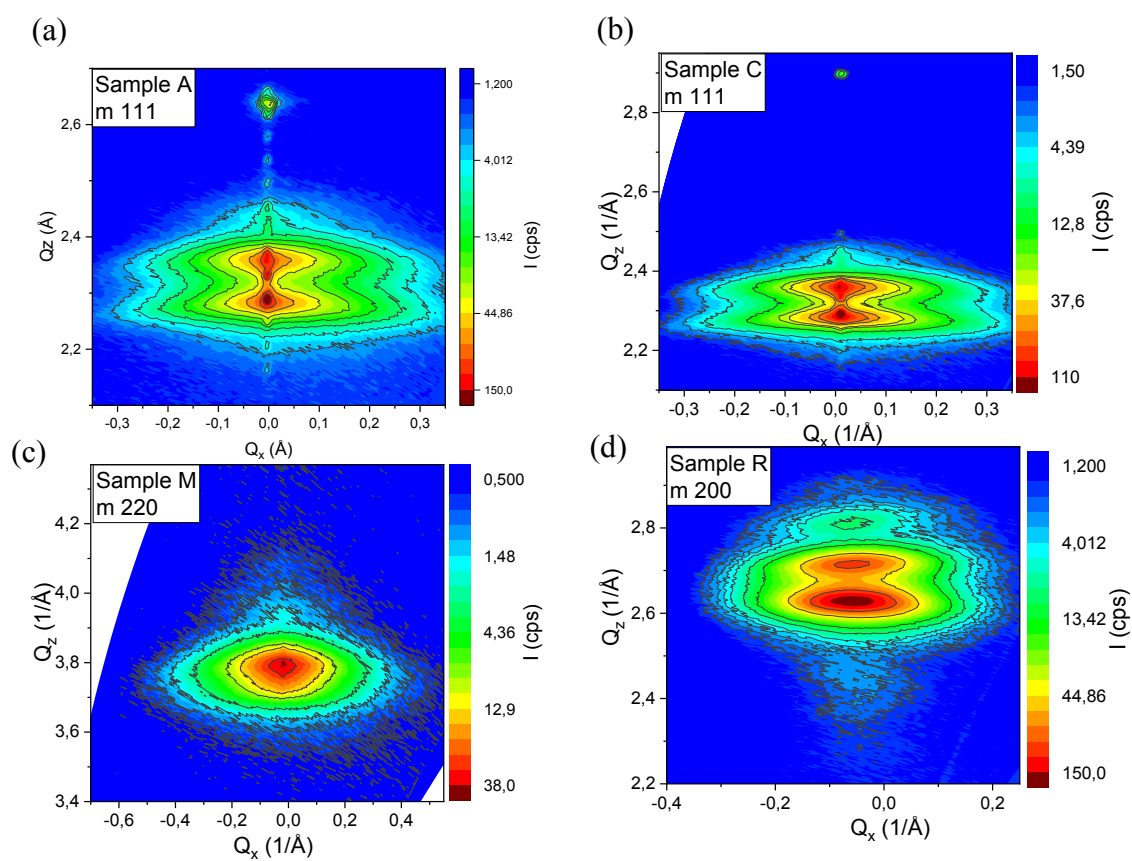

**Figure 4.** High resolution XRD reciprocal space maps of symmetrical 111 {CdO/MgO} (a), (b) 220 {CdO/MgO} (c) and 200 {CdO/MgO} (d) reflections of superlattices in samples A,

C, M and R, respectively.

To check the in-plane correlation of lattice parameters of {CdO/MgO} SLs and sapphire substrates the X-ray pole figures were measured. For sample A (Figure 5a) and sample C (Figure 5b) we observe six diffracted peaks from {CdO/MgO} (111) planes. Thus, one should expect a 3-fold symmetry of the [111] direction. However, since six X-ray diffracted signals were registered, it is evidence of existing twins in our SL system. A similar behavior has been reported in the scientific literature for cubic epitaxial layers of CdO, MgO on *a*-plane and *c*-plane sapphire substrates [9], [14]. For sample M, where SL grows on *m*-plane sapphire, the 200 X-ray diffraction pole figure was registered (Figure 4c). In this case, 4 diffracted signals are observed – two of them are coming from 11-20 sapphire reflections and the other two from 200 {CdO/MgO} SL reflections. The presence of only two {CdO/MgO} diffracted peaks in the XRD pole figure is the evidence of absence of twins in {CdO/MgO} SLs grown on *m*-plane sapphire substrate. For sample R the 224 {CdO/MgO} XRD pole figure were registered (Figure 5d). It exhibits 4-fold symmetry of SL structure. Furthermore, we registered outer ring with 8 diffracted signals. Four of them originate from 224 {CdO/MgO} reflections, whereas the other four from 422 {CdO/MgO} reflections. In this case we also do not observe the twins in the SL structure.

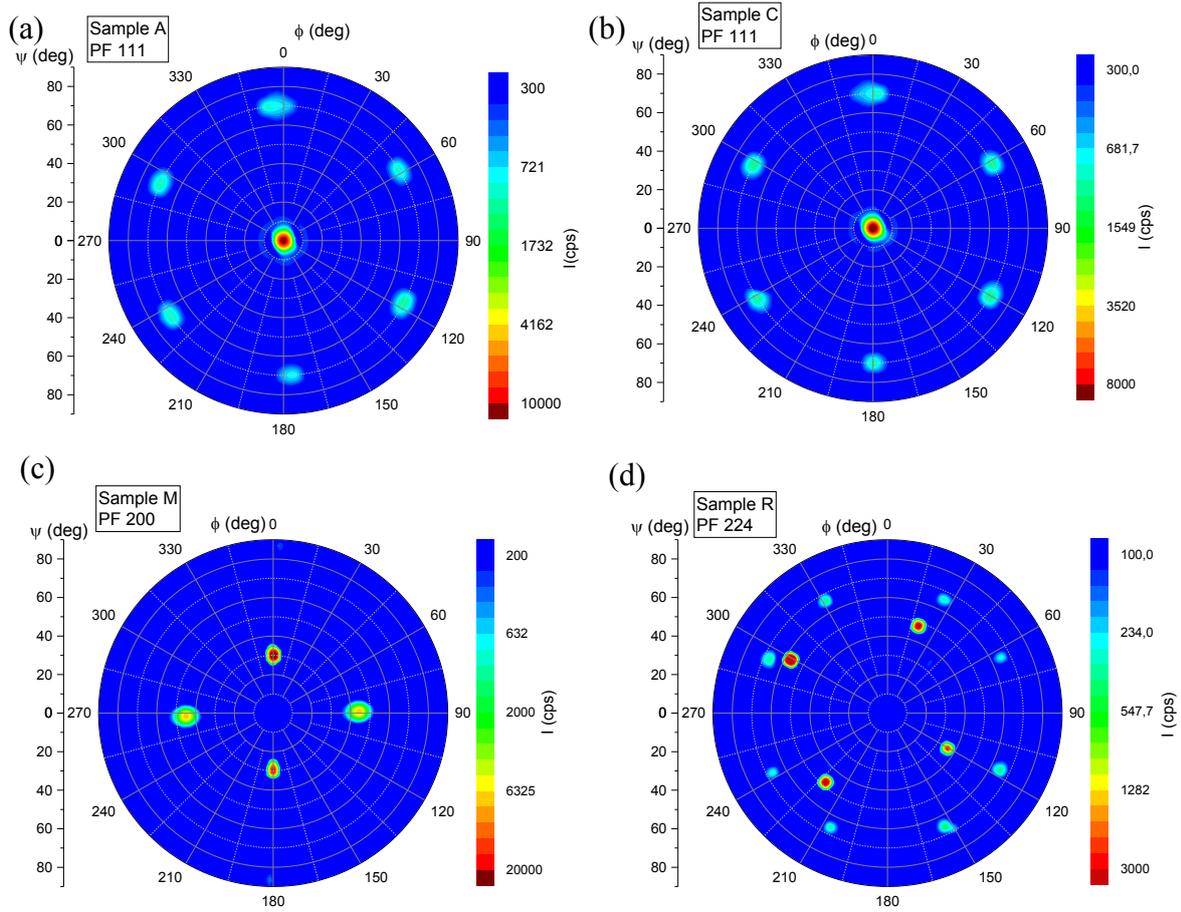

**Figure 5.** XRD pole figures of {CdO/MgO} superlattices on *a*- (a), *c*- (b), *m*- (c) and *r*- (d) plane sapphire substrates.

All of the XRD results were used to create the structural models of a cubic system represented by the {CdO/MgO} SLs on *a*-, *c*-, *m*-, and *r*-plane sapphire substrates. The in-plane epitaxial relationship for $Al_2O_3$ substrate, CdO and MgO is presented in the Figure 6. The substrate's $Al_2O_3$ crystal structure has a rhombohedral R3c cell (space group 167) with lattice parameter $a = b = 4.76$ Å and $c = 12.99$ Å. Crystal structure of α-$Al_2O_3$ has hexagonal symmetry and the structure consists of a hexagonal close-packed array of oxygen atoms with the Al atoms occupying two-thirds of the octahedral interstices. MgO and CdO crystalized in face-centered cubic structure space group Fm-33m (225) cell with lattice parameters $a_{MgO}$ =

$b_{MgO} = c_{MgO} = 4.213$ Å and $a_{CdO} = b_{CdO} = c_{CdO} = 4.695$ Å, respectively. In Figure 6, the oxygen ions (small spheres) have been colored in red, whereas, the aluminum ions (large spheres) are of grey color. The cubic representation of the surface is marked by green and violet lines for CdO and MgO, respectively. Atoms near the interface site in crystal lattices show two possible in-plane twin orientations for (111) cubic MgO or CdO epitaxial crystals on (11-22) Al$_2$O$_3$ substrate. The presence of twins in (111)-oriented rock salt materials grown on *a*-plane sapphire substrates can be visible due to the two-fold symmetry, which exists along [111] CdO or MgO and [11-20] sapphire directions when only one atomic plane on surface is considered. MgO and CdO films have already been grown on *a*-plane sapphire by various methods. The 111 orientation was obtained in case of MgO deposited on *a*-plane sapphire by e-beam evaporation [15]. However, Yadavalli et al. obtained (110)-oriented MgO films on the same substrate material using MBE [16]. In the case CdO obtained by metalorganic vapor phase epitaxy (MOVPE) the 111 orientation was also registered [14]. In the case of the CdO and MgO layers on *a*-plane sapphire, studied by other authors, the presence of twins in XRD pole figures was detected. In our case, twins in SLs on *a*-plane sapphire appeared in the form of six signals in XRD pole figures (Figure 5a) indicating the existence of two sets of crystallites, each rotated by 180°, as it is schematically shown in Figure 6a.

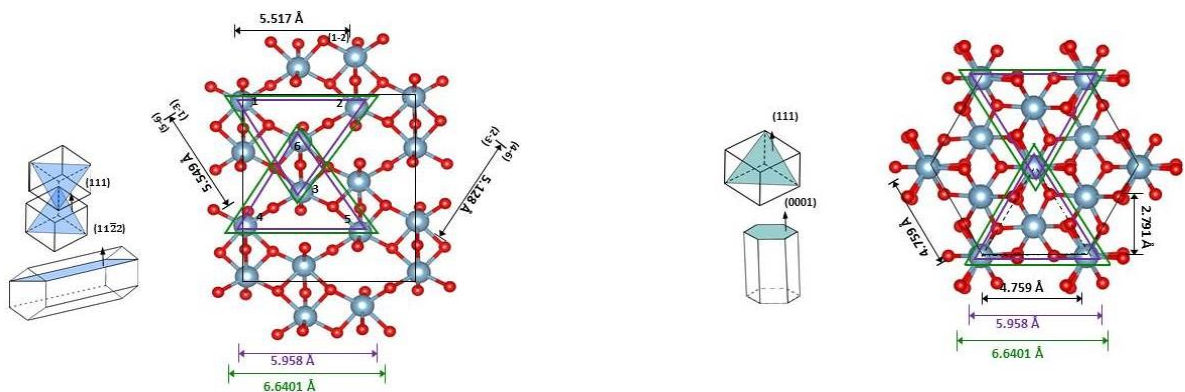
14

(c) (d)

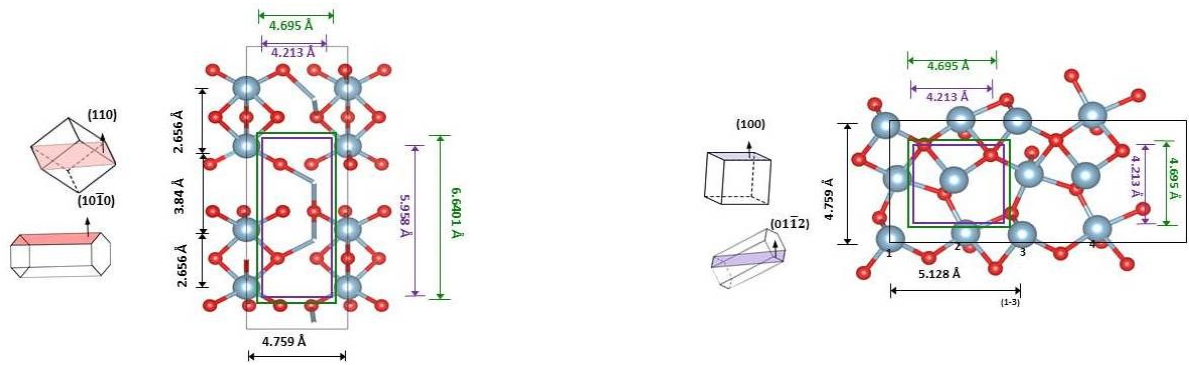

**Figure 6.** Representation of the epitaxial relationship between Al$_2$O$_3$ substrate of different crystallographic orientations and cubic MgO and CdO layers.

As it was mentioned earlier, six XRD signals in the case of {CdO/MgO} SLs on *c*-plane sapphire also revealed the presence of twins. This is because of *c*-plane sapphire lattice surface plane has hexagonal symmetry. Thus, we expect that there may be as many as six sets of crystallites, each rotated in-plane by 60° with respect to each other. The XRD pole figure presented in Figure 5b could be explained by a minimum of two sets of crystallites rotated by 60° relative to each other. It is one of the possible locations of (111) oriented MgO or CdO crystallites on *c*-plane sapphire.

On the (11-20) epitaxial direction of sapphire the {CdO/MgO} SLs grow in the (110) orientation (sample M) with full width at half maximum (FWHM) of rocking curves wider than for the rest of the samples. XRD pole figure for sample M (Figure 5c), determined for 200 {CdO/MgO} reflection, shows only two XRD peaks originating from the SL structure. The other two XRD peaks are assigned to sapphire. Based on the obtained XRD measurement results for sample M a model of the epitaxial relationship between the *m*-plane $Al_2O_3$ substrate and cubic MgO and CdO layers has been proposed and presented in Figure 6c. Figure 6c shows that one unique orientation relationship in the in-plane direction of sample M is possible: it is (110) {CdO/MgO} || (10-10) $Al_2O_3$. The occurrence of only two X-ray diffracted signals proves that twins are absent in the superlattice grown on *m*-plane sapphire (see Fig. 5c).

On *r*-plane sapphire substrate, the {CdO/MgO} SLs grow in (001) crystallographic direction. A model of the relation between the *r*-plane $Al_2O_3$ substrate and the SL in sample R is presented in Figure 6d. It is as follows: (100) {CdO/MgO} || (01-12) $Al_2O_3$. As was already mentioned in the scientific literature the following in-plane relation can be obtained for thin CdO layers on r-plane sapphire: (100) CdO || (2-1-10) Al2O3 and (010) CdO || (01-1-1) Al2O3. Therefore, it can be considered that there is one CdO per two sapphire unit cells along [100] CdO || [2-1-10] Al2O3 crystallographic direction, and three CdO per one sapphire unit cell along [010] CdO || [01-11] Al2O3 [17], [18]. The relatively narrow XRD signals both in the rocking curve and in the pole figure correspond to the four [220] {CdO/MgO} crystallographic diffractions visible in Figure 5d, related to four-fold symmetry confirming a low level of the in-plane disorientation. In this case, number of satellite peaks visible in X-ray scans in Figure 3 (d) and 4 (d) is higher than in the case of other analyzed sapphire substrates, pointing out better crystallographic quality of the structure [9].

Summarizing, the X-ray results obtained for the examined structures showed that the arrangement of atoms on the substrate surface is correlated with its orientation, and has a significant impact on the orientation of the first deposited epi-layers and, consequently, on the orientation of the entire superlattices structure. Furthermore, the symmetry of the substrate and layer also influences the number of expected twin domains [19].

## 2. Raman analysis

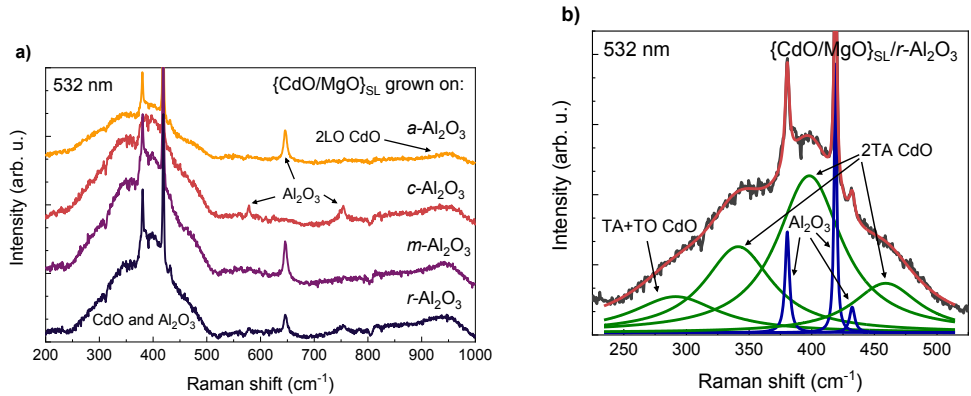

**Figure 7.** (a) Raman spectra of the investigated SL structures grown on $Al_2O_3$ substrates of different crystallographic orientations, (b) the example of elaboration of one of the measured Raman band carried out by fitting the data with Lorentz functions. The laser excitation wavelength was of 532 nm.

Figure 7a presents Raman spectra of the SLs on $Al_2O_3$ substrates of various crystallographic orientations. They are similar to spectra of pure CdO films [20], [21], [22]. Exciting the samples with green laser has caused the manifestation of Raman modes related to CdO and $Al_2O_3$. There are three visible peaks within the range of 550-900 nm that originate from the sapphire substrate. This conclusion has been made by comparing the Raman spectrum of the substrate itself and the spectra of the SL/$Al_2O_3$ structures. Any other $Al_2O_3$ Raman modes have been identified likewise. CdO-related peak of low intensity observed at 950 cm-1 is ascribed to 2LO phonon mode [20], [21]. Furthermore, a complex structure can be observed at 250-550 nm. It consists of several CdO and $Al_2O_3$ modes. A detailed analysis of the spectrum structure is shown in Figure 7b obtained for sample R. Substrate peaks are illustrated as blue curves while the green ones are attributed to the superlattice. Contrary to Raman spectra of CdO [20], [21], [22], in our case the distinguishable TA+TO CdO peak at about 265 cm$^{-1}$ is tough to find. However, when pressure is induced to CdO film, this mode is known to gradually blueshift and lower its intensity [21]. As the examined SLs are strained structures [3], it is possible that it may affect the mode in similar way. Therefore, the broad and weak peak at 291-309 cm$^{-1}$ is identified as TA+TO CdO. The remaining three CdO-related peaks arise from 2LA phonons at different high symmetry points in the Brillouin zone [20]. Data concerning positions of selected modes are collected in Table 2. The broad Raman band at 250-550 nm is also investigated in terms of FWHM of the CdO peaks (see Table 2 as well). Two of four analysed modes in Figure 7a have significantly larger FWHM for *a*-plane sapphire than for other crystal orientations of the substrate. In addition, the {CdO/MgO} SL / *a*-Al$_2$O$_3$ sample is featured by the weakest measured Raman scattering. This finding suggests that the crystal structure quality of the superlattice grown on *a*-Al$_2$O$_3$ is the lowest. Similar results can be found for pure CdO: growing cadmium oxide layer on *a*-Al$_2$O$_3$ leads to extended defects in the film. This does not seem to be an issue for *m*- and *r*-plane sapphire [20].

| Raman Mode | Peak position / FWHM (cm-1) | | | | Excitation wavelength (nm) |
| --- | --- | --- | --- | --- | --- |
| | SL/*a*-Al$_2$O$_3$ | SL/*c*-Al$_2$O$_3$ | SL/*m*-Al$_2$O$_3$ | SL/*r*-Al$_2$O$_3$ | |
| TA+TO CdO | 308 | 303 | 292 | 291 | 532 |
| 2LA CdO | 344/172 | 341/129 | 341/97 | 341/84 | |
| | 397/51 | 403/56 | 398/52 | 398/66 | |

|         | 465/62 | 467/77 | 469/85 | 459/63 |     |
|---------|--------|--------|--------|--------|-----|
| 2TO CdO | 950/108| 949/71 | 948/65 | 948/65 |     |
| TO MgO  | 465    | 453    | 459    | 439    | 320 |

**Table 2.** Raman peaks positions and FWHM values (after the slash) of the observed Raman modes.

Changing the excitation wavelength from 532 nm to 320 nm completely modifies the measured Raman spectra, as shown in Figure 8a. This time, only one Raman mode related to the superlattice structure is visible. Figure 8b presents an exemplary analysis of the obtained data. It appears that CdO-related modes are no longer present in the spectra. The band at 439-471 cm$^{-1}$ can be assigned to TO MgO mode [23]. First-order Raman modes are forbidden in the case of rocksalt compounds [20]. Nevertheless, it is possible to observe them in MgO microcrystals [23], nanoparticles [24], and nanowires [25]. In our structure, the thickness of a single MgO layer is only ca. 2 nm, which may explain the emergence of TO MgO mode. For another rocksalt compound, CdO, inducing high pressure leads to the appearance of disorder-activated first order modes [21]. MgO layers in the superlattice are much more strained than the CdO ones as they are about three times thinner [23]. Thus, disorder introduced by the strain is also a possible reason for their appearance.

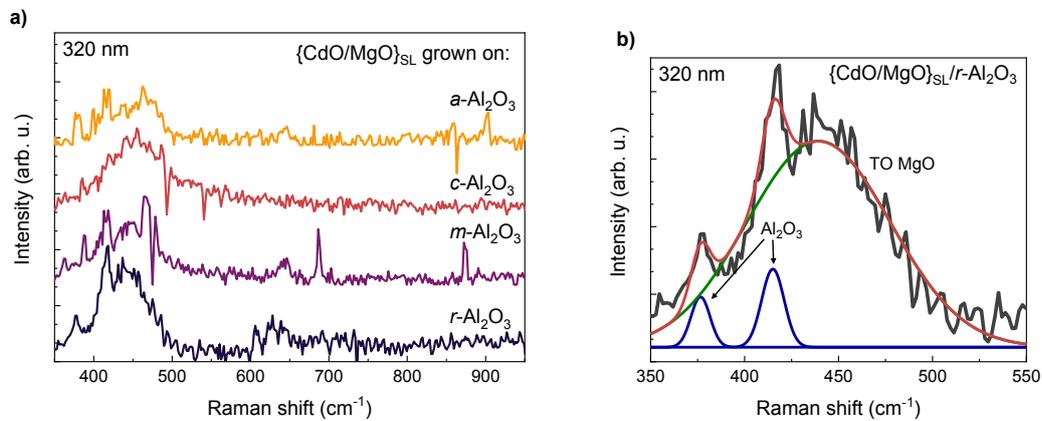

**Figure 8.** (a) Raman spectra obtained for the investigated superlattice structures, (b) an analysis of the selected Raman spectrum range carried out by fitting the data with Lorentz functions. The laser excitation wavelength was of 320 nm.

Position of TO MgO mode fairly varies (see Table 2). One of the reasons is surely its low intensity so it is hard to extract the exact peak position from the data. Similarly to Figure 7a, Raman scattering of SL-related modes is clearly the weakest for the structure grown on *a*-$Al_2O_3$. Additionally, it needs to be mentioned that it is not achievable to determine if, and how, TO MgO mode influences the complex band in Figure 8b.

**Summary**

Structural properties of MBE-grown {CdO/MgO} superlattices on different crystallographic orientations (*a*-, *c*-, *r*- and *m*-plane) of the sapphire substrate have been studied and analyzed by X-ray diffraction and Raman spectroscopy.

HRXRD studies revealed that surface symmetry in both substrates and epi-layers affects the number of twins in the superlattices. No twins and thus better interface quality and better crystallographic quality were obtained for SLs grown on *r*-oriented sapphire. We showed that the growth rate strongly depends on the substrate orientation. For the c-plane and m-plane sapphire substrates we observed thicker CdO sublayers in {CdO/MgO} SLs. Therefore, it can be concluded that the growth rate was comparable for these two substrate orientations. In contrast, the *a*-plane and *r*-plane orientation of the sapphire substrate seemed to favor a slower growth rate of CdO sublayers.

Calculations of the lattice parameters for {CdO/MgO} SLs showed that the lattice constant for sample M differed significantly from the other ones. It is the evidence of lattice distortion existence in {CdO/MgO} SL grown on m-plane sapphire. A similar behavior is observed in Raman spectra.

The best structural quality of {CdO/MgO} SL was achieved for sample R. HRXRD showed that for *r*-plane sapphire substrate orientation the growth is easier to implement. The same conclusion can be drawn based on Raman spectra performed with laser excitation of 325 nm, which showed that the most intense signal from the SL was achieved for the sample grown on *r*-oriented sapphire. In the case of {CdO/MgO} SLs grown both on *a*- and *c*-oriented sapphire substrates the presence of twins in the SLs was confirmed in the XRD experiments. Raman scattering studies revealed the worst crystallographic quality for the sample grown on *a*-plane oriented substrate.

In summary, the heteroepitaxial growth of {CdO/MgO} superlattices on different crystallographic orientations of sapphire through molecular beam epitaxy offers precise control over strain, interface properties, and device compatibility. This control is crucial in optimizing the performance of devices based on these superlattices, making heteroepitaxy an essential technique in their preparation.

**Acknowledgements**

This work was supported in part by the Polish National Science Center, Grants No. 2021/41/B/ST5/00216, as well as the statutory grant (No. 8211104160) of Department of Quantum Technologies of Wroclaw University of Science and Technology.